\documentclass{ws-procs9x6}

\def\beq{\vspace{-0.3em}\color{magenta}\begin{eqnarray*}}
\def\eeq{\end{eqnarray*}\color{blue}\vspace{-0.3em}}
\def\bea{\begin{eqnarray*}}
\def\eea{\end{eqnarray*}}

\def\lQ{\Lambda_{\rm QCD}}

\def\als{\alpha_{\rm s}}

\def\siml{{\ \lower-1.2pt\vbox{\hbox{\rlap{$<$}\lower6pt\vbox{\hbox{$\sim$}}}}\ }}     
\def\simg{{\
    \lower-1.2pt\vbox{\hbox{\rlap{$>$}\lower6pt\vbox{\hbox{$\sim$}}}}\ }}

\newcommand{\be}{\begin{equation}}
\newcommand{\ee}{\end{equation}}

\begin{document}

\title{SYSTEMS of TWO HEAVY QUARKS with EFFECTIVE FIELD THEORIES}

\author{N. BRAMBILLA}

\address{Dipartimento di Fisica dell'Universit\`a di Milano and INFN
\\
Milano, Italy\\
$^*$E-mail: nora.brambilla@mi.infn.it\\}

\begin{abstract}
I discuss results and applications of QCD nonrelativistic effective 
field theories for systems with two heavy quarks.
\end{abstract}

\keywords{QCD, Effective Field Theories, Heavy Quarks}

\bodymatter
\vspace{-1.5mm}
\section{Introduction}\label{sec1}
Systems made by to heavy quarks play an important role in several 
high energy experiments. The diversity, quantity 
and accuracy of the data currently being collected is impressive 
and includes:
data on quarkonium formation from BES at BEPC, E835 at Fermilab,
KEDR (upgraded) at VEPP-4M, and CLEO-III, CLEO-c; 
clean samples of charmonia produced in B-decays, in photon-photon
fusion and in initial state radiation, from the B-meson factory
experiments, BaBar at SLAC and Belle at KEK, including the unexpected
observation of large amounts  of associated $(c\overline{c})(c\overline{c})$ production;
the CDF and D0 experiments at Fermilab measuring heavy quarkonia
production from gluon-gluon fusion in $p\bar{p}$ annihilations at
2~TeV; the Selex experiment at Fermilab with  the preliminary observation of 
possible candidates for doubly charmed baryons; 
ZEUS and H1, at DESY, studying charmonia production in photon-gluon
fusion; PHENIX and STAR, at RHIC, and NA60, at CERN, studying charmonia
production, and suppression, in heavy-ion collisions.
In the near future, even larger data samples are expected from the
BES-III upgraded experiment, while the B factories and the
Fermilab Tevatron  will continue to supply
valuable data for several years.  Later on, new facilities will become
operational (LHC at CERN, Panda at GSI, hopefully a Super-B factory,
a Linear Collider, etc.) offering fantastic challenges and opportunities in this field.  
A  comprehensive review of the experimental and theoretical 
status of heavy quarkonium physics may be found in the Cern Yellow Report prepared by 
the Quarkonium Working Group 
\cite{qwg}. See also the talks at the last QWG meeting at BNL (cf. http://www.qwg.to.infn.it/).

On the theory side, systems made by two heavy quarks are a rather unique laboratory.
They are characterized by the existence of a hierarchy of energy scales in correspondence 
of which one can define a hierarchy of nonrelativistic effective field theries (EFT),
each EFT has less degrees of freedom left dynamical and is simpler. Some of these 
physical scales are large and  may be treated in perturbation theory. 
The occurrence of these two facts makes two heavy quark systems accessible in QCD. 
In particular the factorization of high and low energy scales realized 
in the EFTs allows us to study low energy QCD effects in a systematic and under control 
way. Ultimately the simplest EFT defined at the ultrasoft energy, pNRQCD, allows us to study 
the  Quantum Mechanics of a non-Abelian field theory.

\section{Scales and EFTs}
The description of hadrons containing 
two heavy quarks is a rather challenging problem,  
which adds to the complications of the bound state in field theory 
those coming from a nonperturbative QCD low-energy dynamics.
A  simplification is provided  by  the nonrelativistic nature 
suggested by the large mass of the heavy quarks
and  manifest in the spectrum pattern.
Systems made by two heavy quarks are thus characterized
 by three energy scales, hierarchically ordered by the quark velocity 
$v \ll 1$: the mass $m$ (hard scale), the momentum transfer $mv$ (soft scale), 
which is proportional to the inverse of the typical size of the system $r$,
 and  the binding energy $mv^2$ (ultrasoft scale), which is proportional to the 
inverse of the typical 
time of the system. In bottomonium $v^2 \sim 0.1$, in charmonium $v^2 \sim 0.3$.
In perturbation theory $v\sim \als$. Feynman diagrams will get contributions 
from all momentum regions associated with these scales.
Since these momentum regions depend on $\als$, each Feynman diagram contributes 
to a given observable with a series in $\als$ and a non trivial counting.
For energy scales close to $\lQ$, the scale at which 
nonperturbative effects become dominant, perturbation theory breaks down 
and one has to rely on nonperturbative methods.
 Regardless of this, the non-relativistic hierarchy 
$ m \gg mv \gg mv^2 $ will persist  also below  the  $\lQ$  threshold.

The wide span of energy scales involved makes also a lattice calculation in full QCD 
extremely challenging. 
We may, however, take advantage of the existence of a hierarchy of scales  
by substituting QCD with simpler but equivalent EFTs.
A hierarchy of EFTs may be constructed by systematically integrating out 
modes associated to  energy scales not relevant for the two quark  system.
Such integration  is made  in a matching procedure that 
enforces the equivalence between QCD and the EFT at a given 
order of the expansion in $v$
and achieves a  factorization between the high energy and the low energy contributions.
By integrating out the hard modes one  obtains Nonrelativistic QCD. In such EFT soft 
and ultrasoft scales are left dynamical and still their entanglement complicates 
calculation and power counting. We will focus here on the simplest EFT you can write 
down for two heavy quark systems, where only ultrasoft degrees of freedom remain dynamical.
This is potential NRQCD \footnote{for an alternative and equivalent EFT (in the case in which $\lQ$ is 
the smallest scale) see \cite{vnrqcd}. }.

\section{pNRQCD}
pNRQCD \cite{pnrqcd1,pnrqcd2,reveft} is the EFT for two heavy quark systems that follows from NRQCD 
by integrating out the soft scale. Here the role of the potential and the
 quantum mechanical nature of the problem are realized in  the fact that the Schr\"odinger 
equation appears as zero problem for two quark states.
We may distinguish two situations:
1) weakly coupled pNRQCD when $mv  \gg \lQ$, 
where the matching from NRQCD to pNRQCD may be performed in
 perturbation theory;
2) strongly coupled pNRQCD when $ mv \sim \lQ$, 
where the matching has to be nonperturbative.
Recalling that $r^{-1}
  \sim mv$, these two situations correspond  to systems with inverse
typical radius smaller than or  of the same order as  $\Lambda_{\rm QCD}$.

\subsection{Weakly coupled pNRQCD}
The effective degrees of
freedom are: low energy $Q\bar{Q}$ states (that can be decomposed into  a singlet
and an octet field under colour transformations) 
with energy of order  $ \Lambda_{\rm QCD},mv^2$ and
    momentum   ${\bf p}$ of order $mv$,  
plus  ultrasoft gluons
with energy  and momentum of order
$\lQ,mv^2$. 
All the  gluon fields are multipole  expanded (i.e.  expanded  in
$r$). 
The Lagrangian is then given by terms of the type
\begin{equation}
{c_k(m, \mu) \over m^k}  \times  V_n(r
\mu^\prime, r\mu)   \times  O_n(\mu^\prime, mv^2, \lQ)\; r^n  .
\end{equation}
where the potential matching 
coefficients $V_n$  encode the non-analytic behaviour in $r$.
At  leading order in the multipole expansion,
the singlet sector of the Lagrangian gives rise to equations of motion of the 
Schr\"odinger type. 
Each term in  the pNRQCD Lagrangian has a definite power counting;
retardation (or non-potential) effects 
start at the NLO in the multipole expansion and are systematically 
encoded in the theory, they are typically related to nonperturbative effects
\cite{pnrqcd2,reveft}.
Resummation of 
 large logs (typically logs of the ratio of energy and momentum scales) can 
be obtained
 using the renormalization group  (RG) adapted to the case of correlated scales \cite{rg};
 Poincar\'e invariance is not lost, but shows up 
in some exact relations among the matching coefficients \cite{poincare0}.
The renormalon subtraction 
may be implemented systematically. 
\cite{reveft} 
\vskip 0.05truecm
\leftline{\bf Applications of weakly coupled pNRQCD}
\noindent
{\bf QCD Singlet Static potential.}
The singlet and octet potentials are well defined matching coefficients 
to be calculated in the perturbative matching. In \cite{US}
a determination
of the singlet potential at three loops leading log has been obtained  
and correspondingly  also a determination of $\alpha_V$ showing how 
this quantity starts to depend on the infrared behaviour  of the theory at three loops.
The perturbative calculation of the static potential at (almost) three loops and with 
the RG improvement has been compared to the lattice calculation of the potential and found 
in good agreement up to about 0.25 fm \cite{pertlat}.
\par\noindent
{\bf $b$ and $c$ masses.} 
\par\noindent
Heavy quarkonium is one of the most suitable system to extract a precise 
determination of the mass of the heavy quarks $b$ and $c$.
Perturbative determinations of the 
$\Upsilon(1S)$ and $J/\psi$ masses have been used to extract the $b$ and $c$
masses. The main uncertainty in these determinations
comes from nonperturbative contributions (local and nonlocal condensates \cite{pnrqcd2})
together with possible effects due to
subleading renormalons \cite{reveft}. 
A recent analysis  performed by the QWG \cite{qwg}
and based on all the previous determinations indicates
an error of about 50 MeV both for the  bottom 
($1\% $ error) and in the charm ($4 \% $ error) mass.
\par\noindent
{\bf Perturbative quarkonium  spectrum.}\par \noindent
{\bf $B_c$ mass}. Table \ref{TabBc} shows some 
recent determinations of the $B_c$ mass in perturbation theory at NNLO 
accuracy compared with a  recent 
lattice study \cite{Allison:2004be} and the value of the CDF  experimental  $B_c$ mass. 
This would support 
the assumption that nonperturbative contributions to the quarkonium ground state are of the 
same magnitude as NNLO or even NNNLO corrections, which would be consistent with a 
$ mv^2 \simg\lQ$ power counting.
{\bf Hyperfine splittings.}
$c\bar{c}$, $b \bar{b}$, $B_c$ ground state hyperfine splittings have been recently calculated 
at NLL in \cite{Kniehl:2003ap}. The prediction for $\eta_b$ mass is
$M(\eta_b) = 9421 \pm 10\,{(\rm th)} \,{}^{+9}_{-8}\, (\delta\als)~{\rm MeV}$.
The logs resummation seems to be important.
If the experimental error in future measurements of $M(\eta_b)$
  will not exceed few Mev, the bottomonium hyperfine separation will become a 
  competitive source of $\alpha_s(M_Z)$ with an estimated accuracy  of 
  $\pm 0.003$.
\par\noindent
{\bf Radiative transitions (M1).}
A theory of M1 transitions in heavy quarkonium has been recently formulated
using pNRQCD \cite{M1}. 
This  may shed some light on recent CLEO results on radiative M1 transitions
in the $\eta_b$ search that have ruled out several models. No large anomalous 
quarkonium magnetic moment is generated.
\par\noindent
{\bf Seminclusive radiative decays of $\Upsilon(1S)$.} 
In \cite{radup} the end-point region of the photon spectrum in semi-inclusive radiative decays of 
 heavy quarkonium has been discussed using Soft-Collinear Effective Theory and pNRQCD.
Including the octet contributions a good understanding of the experimental  data is obtained. 
\par\noindent
\par\noindent
{\bf Gluelump spectrum. }
In pNRQCD \cite{pnrqcd2,Bali:2003jq}  the full structure of the gluelump spectrum has been studied,
obtaining model independent predictions on the shape, the pattern, the degeneracy 
and the multiplet structure of the hybrid static energies for small $Q\bar{Q}$ 
distances that well match and interpret the existing lattice data.  
\par\noindent
{\bf Properties of baryons made of two or three heavy quarks.}
Recently the SELEX experiment has detected first signals from three-body bound states made 
 of two heavy quarks and a light one. The two heavy quark part of such systems may be treated 
in pNRQCD and a precise predication for the hyperfine interaction may be obtained \cite{3qeft}.
%


\begin{table}[ht]
\begin{tabular}{|c|cccc|}
\hline
\multicolumn{5}{|c|}{$B_c$ mass ~(MeV)}\\
\hline
\cite{bcexp} (expt) &\cite{Allison:2004be} (lattice) & \cite{Brambilla:2000db} (NNLO)
& \cite{Brambilla:2001fw} (NNLO)& \cite{Brambilla:2001qk} (NNLO)\\
\hline
$6287\pm 4.8 \pm 1.1 $ &$6304\pm12^{+12}_{-0}$ & 6326(29) & 6324(22) & 6307(17) 
\\
\hline
\end{tabular}
\caption{Different perturbative determinations of the $B_c$ mass compared with the experimental 
value and a recent lattice determination.}
\label{TabBc}
\end{table}

\vspace{-10mm}

\subsection{Strongly coupled pNRQCD}
In this case the 
matching to pNRQCD is nonperturbative \cite{pnrqcdnonpert}.
In the situation where  the other degrees of freedom 
(like those associated 
with heavy-light meson pair  threshold production and heavy hybrids) 
develop a mass gap of order $\lQ$, 
the quarkonium singlet field $\rm S$ remains as the only low energy dynamical 
degree of freedom in the pNRQCD Lagrangian (if no ultrasoft pions are considered), 
which reads \cite{pnrqcdnonpert,sw,pnrqcd2,reveft}:
\begin{equation}
\quad  {L}_{\rm pNRQCD}= {\rm Tr} \,\Big\{ {\rm S}^\dagger
   \left(i\partial_0-{{\bf p}^2 \over
 2m}-V_S(r)\right){\rm S}  \Big \} .
\end{equation}
In this regime   we recover the quark potential singlet model from
 pNRQCD.  The matching potential $V_S$ (static and relativistic corrections)
  is nonperturbative: the real part controls
the spectrum and the imaginary part controls the inclusive decays. The potential is  calculated 
in the nonperturbative matching procedure between NRQCD and pNRQCD \cite{pnrqcdnonpert,reveft}.
Advantages of this approach include:
factorization of  hard (in the NRQCD matching coefficients) and  soft  scales
 (contained in Wilson loops or nonlocal gluon correlators);
the  low energy objects being only  glue dependent,
confinement investigations,
        on the lattice 
and in QCD vacuum models become feasible \cite{rev};
the  existence of a   power counting indicating leading and subleading terms 
in quantum-mechanical 
perturbation theory;
the quantum mechanical  divergences (like the ones coming from iterations
    of spin delta potentials)  are absorbed  by NRQCD matching coefficients.
The potentials evaluated on the lattice once used in the 
Schr\"odinger equation produce the spectrum.
The calculations involve only   QCD parameters  (at some scale and in some 
scheme). 

\leftline{\bf Applications of strongly coupled  pNRQCD}
\noindent
{\bf Nonperturbative potentials and Spectrum.}
Recently the multilevel algorithm has been applied to the lattice evaluation 
of field strength insertion inside the Wilson loop average, producing 
very precise data for the spin dependent potentials  and a first evaluation of the nonperturbative potential 
at order $1/m$ \cite{koma}. This is the first step towards a precise determination of 
the nonperturbative matching potentials on the lattice.
\par\noindent
{\bf Decays.}
The inclusive quarkonium decay widths  
in pNRQCD can be  factorized  with respect to the wave function 
(or its derivatives) calculated in zero,
which  is suggestive of the early potential models results:
$
  \Gamma ({\rm H}\to{\rm LH}) = F(\als,\lQ) \, \cdot \, \vert \psi (0)\vert^2 .
$
Similar expressions hold for the electromagnetic decays.
However, 
the coefficient $F$ depends here  both on $\als$ and $\lQ$. In particular 
all NRQCD matrix elements, including the octet
ones, can be expressed through pNRQCD as products of  universal 
nonperturbative factors by the squares of the quarkonium wave functions
(or derivatives of it) at the origin. The nonperturbative factors are typically
integral of nonlocal electric or magnetic correlators and thus  
depends on the glue but not on the quarkonium state \cite{sw}. 
Typically $F$ contains both the NRQCD matching coefficients  at the 
hard scale $m$ and the nonperturbative correlators at the low energy scale $\lQ$.
The  nonperturbative correlators, being state independent, are in a
smaller number than the   nonperturbative NRQCD   matrix elements
and thus the predictive power is increased in going from NRQCD to pNRQCD.
Thus,  having fixed the nonperturbative parameters on charmonium decays, 
new model-independent QCD predictions can be obtained  for the bottomonium decay
widths \cite{sw}.
\vspace{-5mm}

\end{document}